\documentclass[sigconf]{acmart}

\usepackage{graphicx}
\usepackage{textcomp}
\usepackage{xcolor}
\usepackage{booktabs,siunitx}
\usepackage{subfig}
\usepackage{subfloat}

\usepackage{amsmath}
\usepackage[ruled,linesnumbered]{algorithm2e}
\usepackage{multirow}

\AtBeginDocument{%
  \providecommand\BibTeX{{%
    \normalfont B\kern-0.5em{\scshape i\kern-0.25em b}\kern-0.8em\TeX}}}

\setcopyright{acmcopyright}
\copyrightyear{2018}
\acmYear{2018}
\acmDOI{XXXXXXX.XXXXXXX}

\acmConference[Conference acronym 'XX]{Make sure to enter the correct
  conference title from your rights confirmation emai}{June 03--05,
  2018}{Woodstock, NY}
\acmPrice{15.00}
\acmISBN{978-1-4503-XXXX-X/18/06}

\begin{document}

\title{Expected Transaction Value Optimization for Precise Marketing in FinTech Platforms}


\author{Yunpeng Weng}
\email{edwinweng@tencent.com}
\orcid{0000-0001-7593-2169}
\affiliation{%
  \institution{FiT, Tencent}
  \city{Shenzhen}
  \country{China}
}

\author{Xing Tang}
\email{xing.tang@hotmail.com}
\orcid{0000-0003-4360-0754}
\affiliation{%
  \institution{FiT,Tencent}
  \city{Shenzhen}
  \country{China}
}

\author{Liang Chen}
\email{leocchen@tencent.com}
\affiliation{
  \institution{FiT,Tencent}
  \city{Shenzhen}
  \country{China}
}

\author{Dugang Liu}
\email{dugang.ldg@gmail.com}
\affiliation{
  \institution{Guangdong Laboratory of Artificial Intelligence and Digital Economy (SZ)}
  \city{Shenzhen}
  \country{China}
}
 
\author{Xiuqiang He}
\authornote{Corresponding author.}
\email{xiuqianghe@tencent.com}
\affiliation{
  \institution{FiT, Tencent}
  \city{Shenzhen}
  \country{China}
}


\begin{abstract}
	
FinTech platforms facilitated by digital payments are watching growth rapidly, which enable the distribution of mutual funds personalized to individual investors via mobile Apps. As the important intermediation of financial products investment, these platforms distribute thousands of mutual funds obtaining impressions under guaranteed delivery (GD) strategy required by fund companies. Driven by the profit from fund purchases of users, the platform aims to maximize each transaction amount of customers by promoting mutual funds to these investors who will be interested in. Different from the conversions in traditional advertising or e-commerce recommendations, the investment amount in each purchase varies greatly even for the same financial product, which provides a significant challenge for the promotion recommendation of mutual funds. In addition to predicting the click-through rate (CTR) or the conversion rate (CVR) as in traditional recommendations, it is essential for FinTech platforms to estimate the customers' purchase amount for each delivered fund and achieve an effective allocation of impressions based on the predicted results to optimize the total expected transaction value (ETV). In this paper, we propose an ETV-optimized customer allocation framework (EOCA) that aims to maximize the total ETV of recommended funds, under the constraints of GD dealt with fund companies. EOCA consists of two phases: a prediction phase of the customer purchase amount followed by a constrained allocation phase. Specifically, we propose an entire space deep probabilistic model with a novel-designed loss function to predict the purchase amount when a promotional fund is exposed to a user, which involves not only the conversion rate prediction but also the post-conversion purchase amount estimation. Based on the predicted ETV, we design a heuristic algorithm to solve the large-scale constrained combinatorial optimization problem to suggest which fund each user should be exposed to in order to maximize the total purchase amount. To the best of our knowledge, it's the first attempt to solve the GD problem for financial product promotions based on customer purchase amount prediction. We conduct extensive experiments on large-scale real-world datasets and online tests based on LiCaiTong, Tencent’s wealth management platform, to demonstrate the effectiveness of our proposed EOCA framework.
\end{abstract}





\maketitle
\section{Introduction}
There are more than 600 million people who invest funds in China \footnote{https://www.amac.org.cn/researchstatistics/report/tzzbg}. With the emergence of digital payments, Chinese individual investors prefer purchasing mutual funds on Internet platforms instead of traditional financial institutions. Therefore, FinTech platforms that cooperate with fund companies for mutual fund distribution via mobile Apps are developing rapidly such as TianTian,  Ant Group, and Tencent LiCaiTong \cite{hong2019fintech}. 
\begin{figure}[h]
	\centering
	\includegraphics[width=0.9\linewidth]{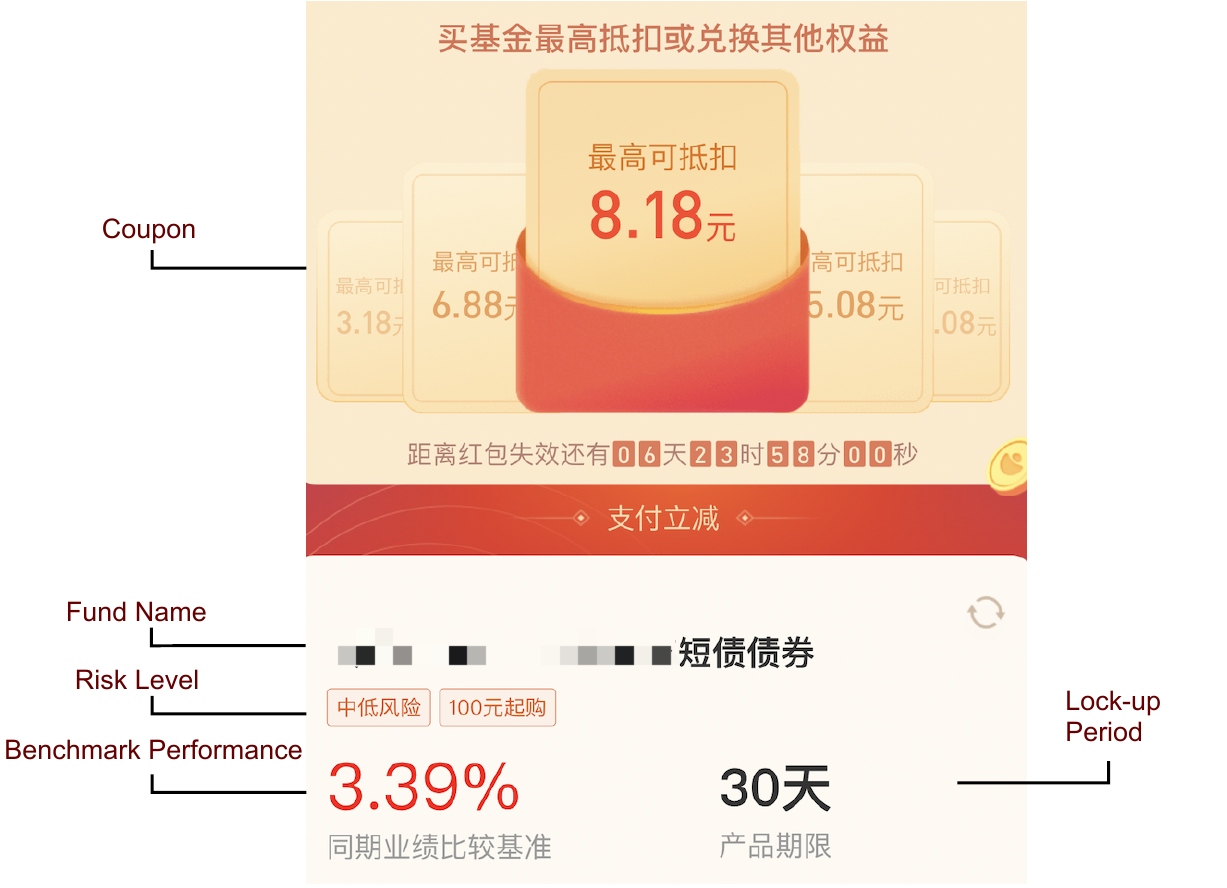}
	\caption{An example of promotional delivery operation on an Internet FinTech plaform.}
	\label{figure_pro}
\end{figure}

Data-driven customer allocation is a widely used growth hacking manner in FinTech platforms for user growth or customer acquisition operations. Generally, financial marketing operations require to allocate customers to different fund type preference groups according to their profiles and historical behavior. FinTech platforms continuously deliver promotional operations about a certain type of funds to the allocated users for precise marketing and aim to boost the platform's AUM.  Figure \ref{figure_pro} provides an example of promotion in a financial distribution platform. 
Due to commercial considerations and the requirements of fund companies, each fund type should obtain a certain number of users with a higher risk tolerance than the risk level of the corresponding financial product type. In other words, for $k$ types of financial products such as Bond Funds and Money Funds, users need to be divided into k groups, and the number of users in each group is predetermined. This leads to a large-scale allocation problem that requires the allocation framework to maximize the total transaction value under multiple constraints. To achieve the maximum transaction value, i.e. the total customers' purchase amount, it's essential for the allocation framework to accurately estimate the customers' expected transaction value (ETV) for each type of fund. In the FinTech platform, since users can freely determine the amount of investment, the transaction value of each successful conversion varies greatly even for the same fund. Nevertheless, for conventional e-commerce or online advertising recommendation,  the expected revenue of a single item display can be calculated only by accurately predicting the click rate/conversion rate since the price of the product or the bid of the advertiser are predetermined. Meanwhile, little literature has studied the problem of predicting customer purchase amount for industrial applications. Although there have been some efforts that use deep learning models for financial product recommendation based on CTR/CVR prediction \cite{DBLP:journals/corr/abs-2008-02546,zheng2021graph}, we argue that the CVR-based allocation strategies may result in sub-optimal total ETV since they don't optimize the ETV directly. It is non-trivial to predict the transaction amount in financial commerce due to the significant zero value inflation and high variance of the transaction amount.  Despite Mean Squared Error (MSE) has become the de facto standard loss function for continuous value prediction problems, it relies on the homoskedasticity assumption which assumes different samples have the same variance. However, the samples in FinTech platforms are heteroscedastic because the investment intention and available capital vary greatly in regard to different user-fund pairs. Although the previous literature proposed a deep learning method with zero-inflated lognormal (ZILN) loss that accommodates zero-inflated distribution for customer lifetime value (LTV) prediction \cite{wang2019deep},  it is not sufficient to meet the needs of growth marketing in actual business. To give a specific example, given a negative sample, suppose that the estimated conversion rate $p = 0.0001 $ (which is quite accurate), while the wrongly estimated post-conversion purchase amount is a too much high value like 1 million CNY since ZILN does not perform supervised learning on the non-converted sample space in the regression task. It makes the expected transaction value overestimated and leads to bad allocation and recommendation results. Besides, in practical industrial applications, the allocation strategies mainly depend on manual operations \cite{lei2020multi}. For example, operation specialists determine the priority of all fund types and allocate customers to high-priority fund type segments in order.  However, the manual allocation strategy highly relies on human experience and easily leads to the sub-optimal total sales amount.

To address the above challenges, we design a two-stage ETV-optimized customer allocation framework (EOCA). The task of the first stage is predicting the customer purchase amount as the expected transaction value for each fund type and the task of the second stage is allocating the customer to different fund types for marketing operations. To alleviate the problems of MSE and ZILN, we propose a novel designed loss function, entire space multi-task joint (ESJ) loss, that unifies the entire sample probability space for users' purchase intention and expected transaction value modeling. Furthermore,  we propose the entire space deep probabilistic model (ESDPM) for the multi-task learning problem of financial recommendations based on ESJ loss. Then, our framework takes all potential customers' expected transaction value for each fund type as constants and uses an efficient heuristic algorithm to search the near-optimal allocation plan.

The main contributions of our work are summarized as follows:
\begin{itemize}
	\item We design an ETV-optimized customer allocation framework that aims to maximize the total customer purchase amount for promotional recommendations in FinTech platforms under the allocated customer size and risk tolerance constraints.  
	
	\item We propose a novel entire space multi-task joint loss function and an entire space deep probabilistic model to predict the expected transaction value of each financial product's delivered impression.  To our best knowledge, it's the first attempt to estimate customer purchase amount for financial products recommendations. 
	
	\item We design an efficient heuristic algorithm to solve the customer allocation problem based on the predicted ETV, which aims to maximize the total transaction value.
	
	\item Comprehensive offline and online experiments are conducted to demonstrate the effectiveness of our proposed EOCA framework. 
\end{itemize}

The rest of this paper is organized as follows. In section \ref{section3}, we provide related works.  We describe the problem formulation in section \ref{section2}. Our proposed framework is presented in section \ref{section4}.  The offline and online experiments are illustrated in section \ref{section5}. In section \ref{section6}, we present the conclusion.

\section{Related Work}\label{section3}

User preference modeling is essential for recommendation systems or delivery systems to show users the products they are willing to purchase.  For the performance-driven online advertising and e-commerce recommendations, the most widely used methods are focus on predicting users' CTR or CVR to reflect users' estimated response \cite{lu2017practical,li2020hierarchical,zheng2021graph,chen2016deep}.  
In recent years, a lot of deep learning-based CTR/CVR prediction models have been proposed. Ma et al. \cite{ma2018entire} pointed out that the CVR models trained with samples of clicked impressions may encounter performance issues in practice due to the sample selection bias. They proposed a multi-task learning method Entire Space Multi-task Model (ESMM) to estimate CTR and post-click CVR in the entire sample space. Furthermore, Wen et al. \cite{wen2020entire} proposed the Elaborated Entire Space Supervised Multi-task Model ($ESM^2$) derived from ESMM. $ESM^2$ considers several purchase-related actions like adding to wish list besids click behavior to alleviate the data sparsity problem. Chapelle \cite{chapelle2014modeling} suggested taking delayed feedback into consideration for online advertising and proposed the delayed feedback model (DFM), a deep probabilistic model that captures the conversion delay. To better estimate users' responses, a lot of studies have been made that utilize sequential models to capture users' historical behavior patterns \cite{zhou2018deep,xiao2020deep,wu2019session}. Su et al. \cite{su2020attention} proposed an attention-based model that captures users' purchase interest from historical clicks and calibrates the CVR through modeling the users' delayed feedback.

Most user response estimation models are proposed for online advertising, e-commerce or information feeds recommendation. Only a few researches have delved into conversion rate prediction for financial products, let alone the expected financial transaction value prediction. Researchers from Ping An, an insurance company in China, proposed HConvoNet which combines the information of users' static profiles and the conversations between insurance agents and customers to predict users' purchase intention \cite{kang2019heterogeneous}. Huang et al. \cite{DBLP:journals/corr/abs-2008-02546} utilize a graph neural network to capture users' dynamic interests for next-purchase financial product prediction. Zheng et al. \cite{zheng2021graph} proposed the graph-convolved factorization machine (GCFM) to model multiple feature interactions for financial recommender systems, they conducted offline experiments on real-world financial datasets. 

However, the existing literature for user response estimation mainly focuses on the CTR/CVR prediction. To the best of our knowledge, there is few work has discussed the problem of customers' purchase amount prediction for the financial product recommendation. For e-commerce or online advertising, there is no need to estimate the transaction amount to make recommendations for maximizing total revenue. Because the price of the product or the advertiser's bid is known in advance. For a specific instance, for cost-per-conversion (CPA) payment advertising systems, the expected transaction value for per thousand  impression is calculated as $eCPM = Pr(click) \times Pr(conversion|click) \times CPA \times 1000$, where CPA is the price the advertiser pays for each conversion. In this case, recommendation models only need to predict the click and post-click conversion rate accurately for optimizing the total revenue. Pei et al. \cite{pei2019value} suggested that it is essential to estimate the total revenue for recommendation systems. They calculate the overall value of exposure by modeling the connections between the user's arbitrary behavior (i.e. clicking, adding to the cart, adding to wishlists) and conversion rate and combining the price of the candidate item to calculate the expected value of the exposure for ranking. Nevertheless, their method is essentially to estimate the probability of various user behaviors and the conversion rate under corresponding conditions since the price of the product does not need to be estimated. For each conversion, the user's investment amount is arbitrary. Therefore, we need to estimate not only the conversion rate but also the transaction amount after the conversion occurs.
The most similar works study the problem of customer lifetime value (CLTV) prediction that does not involve in item recommendation \cite{chamberlain2017customer,bauer2021improved,chen2018customer}. Most previous works predicted CLTV based on regression models with vanilla MSE loss.  Wang et al. \cite{wang2019deep} pointed that MSE loss is sensitive to extremely large LTV samples and limits the performance due to the large fraction of zero LTV samples. Therefore, they proposed the ZLIN loss which consists of cross-entropy loss for classification and regression loss for non-zero samples. 

Regarding the guaranteed delivery (GD) strategies, a variety of literatures have studied the impression allocation problem for guaranteed display advertising \cite{chen2012ad,bharadwaj2012shale,fang2019large}.  Most of the recent related researches are aimed at performance advertising based on CTR/CVR prediction as well. Lei et al. \cite{lei2020multi} proposed the pv-click-ctr model to predict the CTR trends with respect to each video's impression number. They considered the optimization problem with objectives of maximizing the overall video views volume and fairness and suggested using a genetic algorithm to search for a near-optimal solution. Besides, they deployed their model on youku, which is one of the largest online video service platforms in China, and achieved remarkable performance compared to the manual strategy. Zhang et al. \cite{zhang2020request} designed a request-level guaranteed delivery advertising planning system RAP to improve the delivery rate and play rate. RAP includes impressions forecasting and allocation optimization which involves the consideration of CTR prediction.

\section{Problem Formulation}\label{section2}

 The financial customer allocation problem can be formulated as a combinatorial optimization problem with constraints. Given the user set $U$ with N selected potential users and $k$ types of funds, it requires the allocation system to divide the N users into $k$ disjoint groups. Each type of fund $f_j$ will be displayed to users in the corresponding segment. The size of each segment $S_j$ should exactly equal $d_j$. With the expected transaction value $ETV_{ij}$ which can be estimated by models, our aim is maximum the total expected transaction values by optimazing $X$:

\begin{equation}
	{\underset {X}{\operatorname {arg\,max} }} \sum_{i\le N,j\le K}   X_{i,j} \cdot ETV_{i,j}  \label{optimal} ,
\end{equation}

\begin{equation}
	s.t. \qquad  X_{i,j} \in {\{0,1}\} \label{cons_1},
\end{equation}
\begin{equation}
	\sum_{j\le K} X_{i,j} = 1, \forall i\in \{1,2,\dots ,N\} 	\label{cons_2},
\end{equation}
\begin{equation}
	\sum_{i \le N} X_{i,j} = d_j, \forall  j\in \{1,2,\dots ,K\} \label {cons_3},
\end{equation}
\begin{equation}
	X_{i,j} = 1 \rightarrow t_i \ge r_j ,\forall i\in \{1,2,\dots ,N\}  ,\forall j\in\{1,2,\dots ,K\}. \label{cons_4}
\end{equation}
where constraint (\ref{cons_2}) requires that one and only one type of funds will be delivered to each user and constraint (\ref{cons_3}) ensures each product $f_k$ will be delivered to exact $d_k$ users. In the first stage, the allocation framework adopts trained models to predict the $ETV$ and uses it as a constraint to solve the above optimization problem. Especially, constraint (\ref{cons_4}) guarantee that customers can only be allocated to the customer segments of the fund type with a lower risk level $r$ than their risk tolerance $t$.

\begin{figure}[!t]
	\centering
	\includegraphics[width=1.0\linewidth]{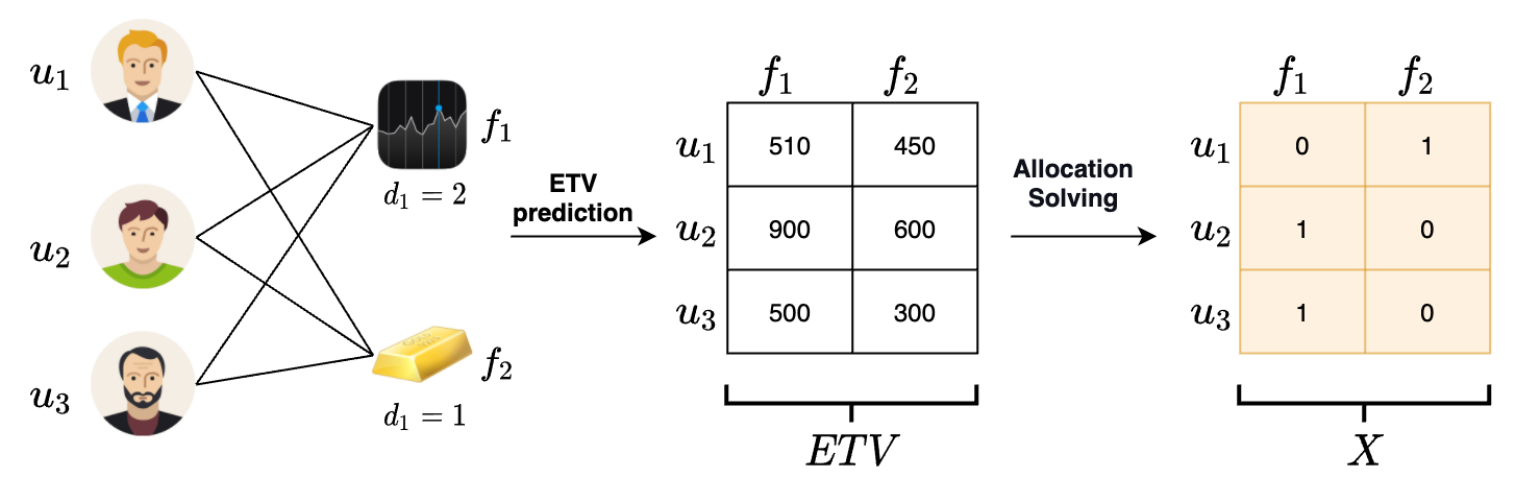}
	\caption{An illustrative example of customer allocation workflow.}
	
	\label{workflow}
\end{figure}

\section{methodology }\label{section4}
 Figure \ref{workflow} presents an illustrative example of customer allocation in the financial distribution platform. In this instance, customers are divided into two segmentations for promotional operations of two different types of funds. Fund type $f_1$ demands two allocated users and fund type $f_2$ is planned to deliver to one user according to the predetermined contract between the FinTech platform and fund company. Firstly, the framework needs to estimate the $ETV_{i,j}$ for each $u_i$ and $f_j$. Secondly, the framework solves the customer allocation problem where the predicted ETV is used as the constant.

\subsection{Entire Space Deep Probabilistic Model}

In order to accurately estimate the ETV, it requires the model to address two main tasks: i) predict whether the user will purchase the delivered item, ii) if the conversion occurs, predict how much money the user is willing to invest.  Thus, we formulate the ETV as the product of click \& conversion rate (ctcvr) and the purchase amount after conversion:
\begin{equation}
	ETV = \underbrace{Pr(click) \times Pr(conversion| click) }_{pCTCVR} \times  \mathbb{E}(PA)
	\label{ETV} 
\end{equation}
where PA denotes the purchase amount(PA) after conversion. Following previous work~\cite{wang2019deep}, we adopt a logarithmic transformation and denote $v$ as the result of the following transformation applied to the $PA$:

\begin{equation}
	v=\left\{
	\begin{aligned}
		&\log (PA + 1)  , &  PA > 0 \\
		&0  , & PA = 0
	\end{aligned}
	\right.
\end{equation}
Note that to avoid confusion between positive samples with an amount of 1 and those without conversion, we added 1 to the purchase amount of all samples, and the processed revenue label is denoted as $y_v = PA + 1$ and thus $v = \log(y_v)$. 

The first task is a typical classification problem which quite a lot of works have discussed. Nevertheless, there are several significant challenges for predicting customer purchase amount and we doubt that the existing methods based on MSE loss or ZILN loss can effectively solve this problem in practice. In addition to the zero inflation phenomenon caused by a large number of non-converted samples, the difference in the purchase amount of converted samples is also very large. 
\begin{figure} \centering    
	\subfloat[The distribution of overall positive samples.] {
		\label{fig:a}
		\includegraphics[width=0.46\columnwidth]{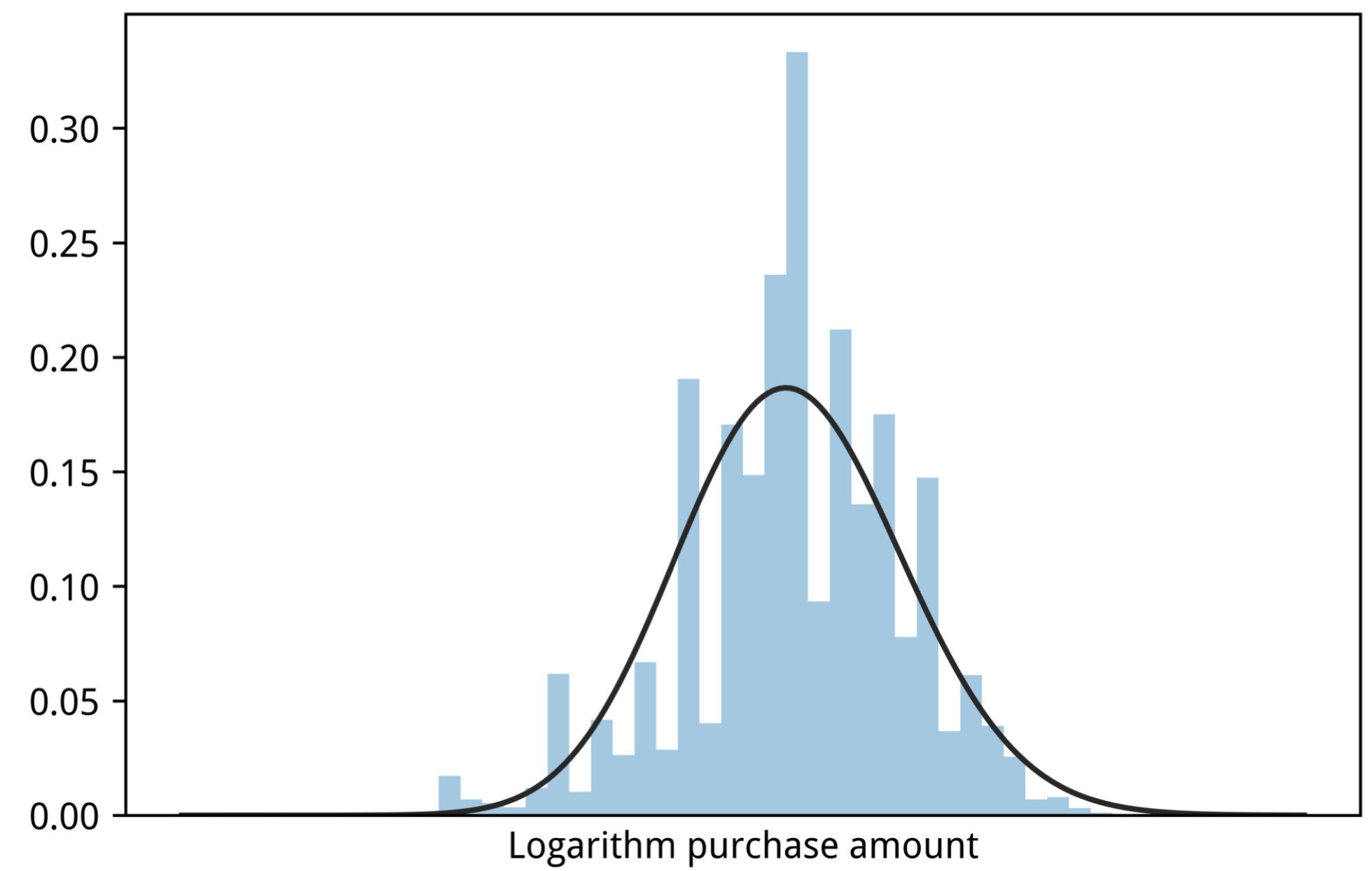}  }     
	\subfloat[Distributions of three users' histroical logarithmic transaction value.] { 
		\label{fig:b}
		\includegraphics[width=0.47\columnwidth]{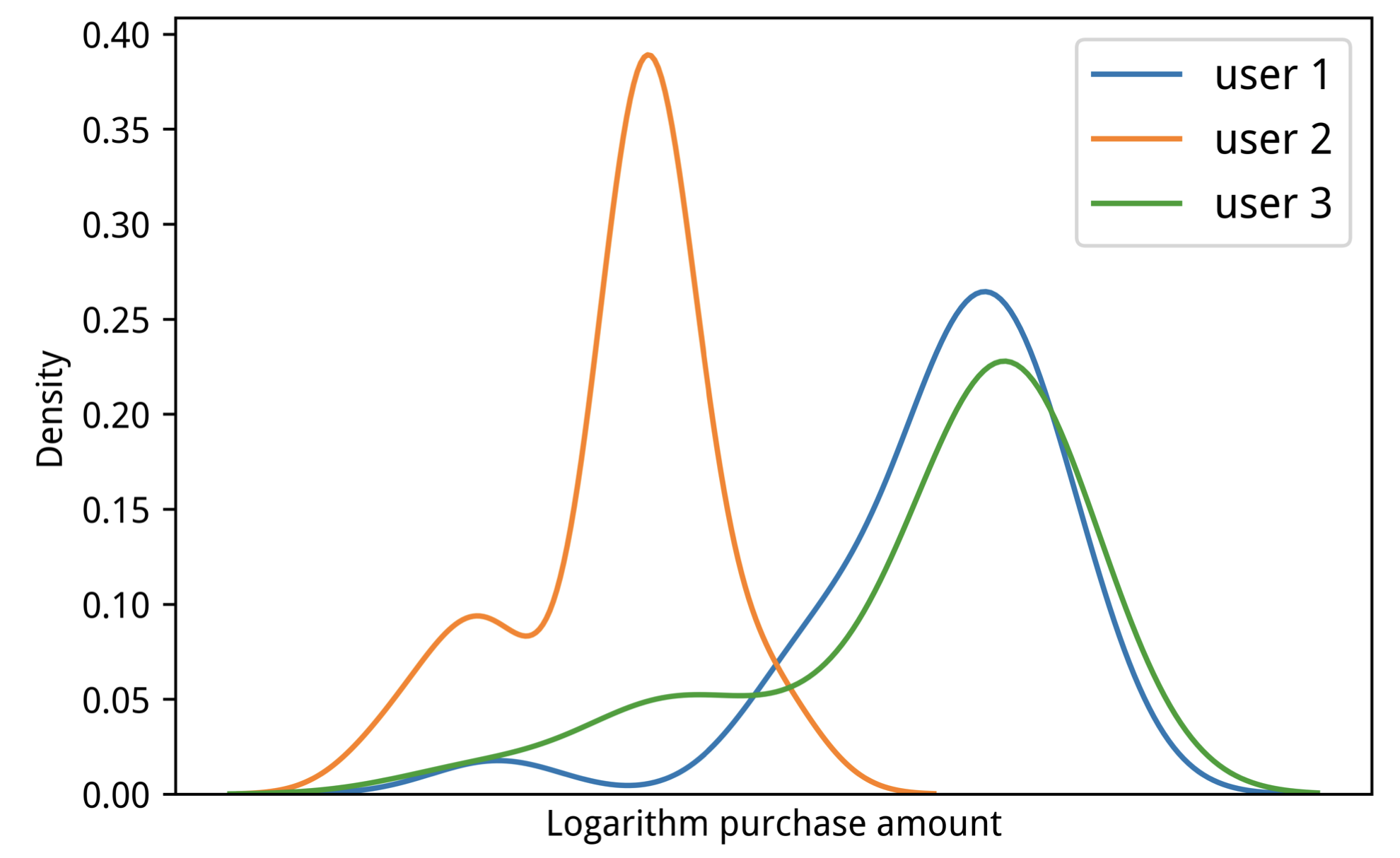}}    
	\caption{ Distribution of logarithmic transaction value. }     
	\label{distribution}     
\end{figure}
Figure \ref{distribution} shows the logarithmic transaction value of converted samples of the real-world financial transactions data. It implies that the converted users' log transaction values follow a normal distribution.  Especially, besides the overall distribution of overall converted samples, we present examples of historical purchase amount distribution of three different users in Figure \ref{fig:b}. User 1 and user 3 have similar profile and behavior patterns while they are quite different from user 2. Meanwhile, the logarithmic transaction value distributions of user 1 and user 3 are similar while the distribution of user 2 is significantly different from theirs. Therefore, inspired by the ZILN loss, we assume that the observed log purchase amount $v_i$ of each positive sample $i$ for a financial product is sampled from an implicit normal distribution. The parameters of the distribution, i.e. the mean $\mu$ and the standard deviation $\sigma$, are determined by the sample's features.

\subsubsection{Entire Sample Space Modeling} For long-term marketing operations, models need to infer both the CTCVR and post-conversion purchase amount on the entire sample space including new coming users. This requires the models are trained with supervision for both tasks on the entire sample space. Otherwise, the performance of models in practical applications will suffer from sample selection bias in the actual industrial environment.

In order to tackle this problem, we provide a novel perspective of sample space division that unifies the classification and regression tasks for financial recommendation scenarios. For the convenience of understanding, we introduce a hidden variable $C$, $C = 1$ means that the user has the willingness to convert for the delivered mutual fund, and $C = 0$ indicates that the user has no intention to purchase. Note that if the conversion is observed means that C = 1 while users might have the intention to convert even for observed negative samples. We assume that the probability of the event $C=1$ is approximately equal to the predicted CTCVR and
\begin{equation}
	\begin{aligned}
		P(v| C = 1, \mu,\sigma) \approx P(v| y = 1, \mu, \sigma) 
		= \frac{1}{\sqrt{2\pi}\sigma y_v }\exp({-\frac{(v- \mu)^2}{2\sigma ^2}})
	\end{aligned}
\end{equation}

\begin{figure*}[h]
	\centering
	\includegraphics[width=1.0\linewidth]{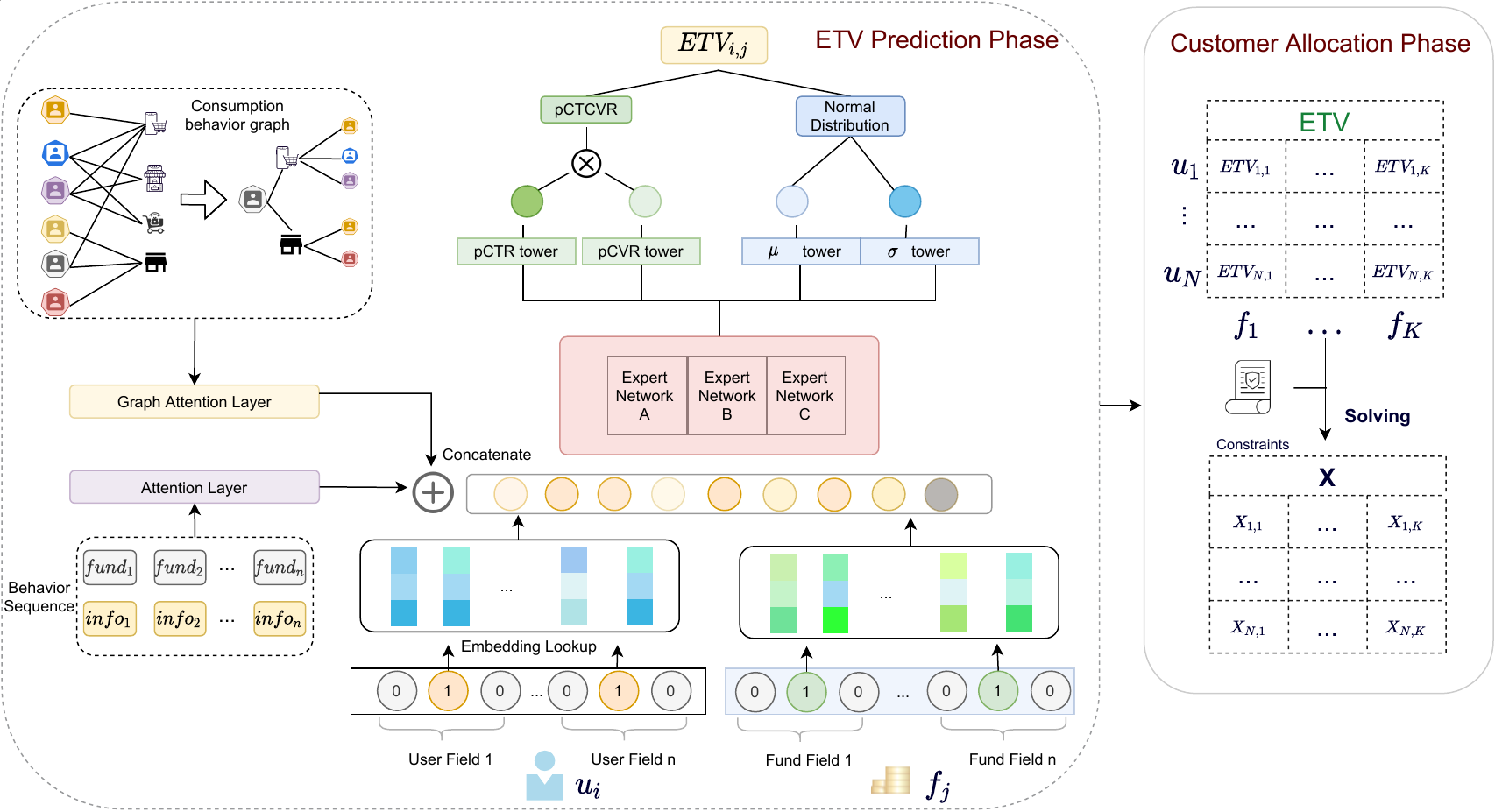}
	\caption{The ETV-optimized customer allocation (EOCA) framework.}
	
	\label{model}
\end{figure*}

For a given data set with each point described as  $(\vec{x_u}, \vec{x_f}, y, v ) $, where $\vec{x_u}$ is features of user $u$ and $ \vec{x_f}$ is features of fund type $f$ such as average annual rate of return, we summarize the sample space as:

	i) User $u$ purchases $f$ and thus $C=1$ and $y = 1$. The observed transaction amount $v \textgreater0$ (Sample subspace $\mathcal{R}_1$).
	ii) User $u$ has no intention to purchase ($C = 0$) and thus $y = 0$ and $v = 0$ (Sample subspace $\mathcal{R}_2$). 
	iii) User $u$ has intention to putchase ($C=1$) while has not purchased yet and thus the observed transaction amount $v=0$ (Sample subspace $\mathcal{R}_3$). 
Samples in the subspace $\mathcal{R}_1$ are observed positive samples and samples in subspaces  $\mathcal{R}_2$ and  $\mathcal{R}_3$ are all observed negative samples. We utilize a multi-task probabilistic model to fit the data set. The first main task is to model the probability of purchase event occurs,  and the second main task is to learn the parameters $\mu$ and standard deviation $\sigma$ of each sample's implicit normal distribution thereby estimate the probability of the event occurs which the sample's observed logarithmic transaction value is $v$ under the condition that C = 1. 

In this way, the probability of observed positive sample with a user-fund pair $i$ that involves user $u$ and fund $f$ can be formulated as:
\begin{equation}
	\begin{aligned}
		& P(C = 1,y = 1, v = v | \vec{x_{u}}, \vec{x_{f}}) \\ 
		&= P(y  = 1 | \vec{x_{u}}, \vec{x_{f}})  \times P(v| \vec{x_{u}}, \vec{x_{f}} , y  = 1) \\
		& = pCTCVR \times \frac{1}{\sqrt{2\pi}\sigma_i y_{vi} }\exp({-\frac{(v_i - \mu_i)^2}{2\sigma_i ^2}})
		\label{positive} 
	\end{aligned}
\end{equation}
The probability of observed negative sample $P(y  = 0, v = 0 | x_{u}, x_{f})$ can be formulated as:
\begin{equation}
	\begin{aligned}
		& P(y = 0, v = 0 | \vec{x_{u}}, \vec{x_{f}}) \\ 
		& = P(C = 0, y = 0, v = 0 | \vec{x_{u}}, \vec{x_{f}}) +   P(C = 1, y = 0, v = 0 | \vec{x_{u}}, \vec{x_{f}}) \\
		& \approx (1 - pCTCVR) + pCTCVR \times \frac{1}{\sqrt{2\pi}\sigma_i }\exp({-\frac{ \mu_i^2}{2\sigma_i ^2}}) 		
		\label{negative} 
	\end{aligned}
\end{equation}
Therefore, the likelihood of ESDPM with trainable parameters set $\Theta$ and data set $\mathcal{D}$ can be written as:

\begin{equation}
	\begin{aligned}
		\mathcal{P}(\mathcal{D};\Theta)   & =  \prod_{ i \in \mathcal{R}_1} P( C = 1, y = 1, v = v| \vec{x_{u_i}},\vec{x_{f_i}}  )   \\
		\times & \prod_{ j \in \mathcal{R}_2 \cup \mathcal{R}_3}  ( P( C = 0, v = 0| \vec{x_{u_j}},\vec{x_{f_j}} ) + P( C = 1, v = 0| \vec{x_{u_j}},\vec{x_{f_j}} ) )
		\label{likehood} 
	\end{aligned}
\end{equation}

According to Equation \ref{positive} - \ref{likehood}, we can finally derive the following form of entire space multi-task joint (ESJ) loss function from the negative log-likelihood loss function: 

\begin{equation}
	\begin{aligned}
		\mathcal{L} & = -\frac{1}{M}(\sum_{ i,y_i = 1 } \log ( P(y = 1, v = v_i | \vec{x_{u_i}}, \vec{x_{f_i}}) ) \\
		& \quad  \quad +\sum_{ j, y_j = 0 } \log(P(y = 0, v = 0 | \vec{x_{u_j}}, \vec{x_{f_j}}) )	\\
		& = -\frac{1}{M}(\sum_{ i,y_i = 1 }( \log p +  \log \frac{1}{\sqrt{2\pi} \sigma_i y_{vi}} +({-\frac{(v_i - \mu_i)^2}{2\sigma_i ^2}})) \\
		&\quad  \quad  + \sum_{j, y_j = 0 }(  \log  (1- p   +  p \times \frac{1}{\sqrt{2\pi}\sigma_j }\exp({-\frac{\mu_j ^2}{2\sigma_j ^2}}))))
		\label{loss function} 
	\end{aligned}
\end{equation}
where we denote pCTCVR as $p$. Through the proposed loss function ESJ, the model has a joint probability modeling of the classification task and regression task on entire sample space. With the learned $pCTCVR_{uf}$ and $\mu_{uf}$ the predicted expected transaction value when fund $f$ is deliverd to user $u$ can be written as:
\begin{equation}
	\begin{aligned}
		{ETV} = pCTCVR_{uf} \times (\exp(\mu_{uf} + \sigma_{uf}^2/2) - 1 )
		\label{final etv} 
	\end{aligned}
\end{equation}

\subsubsection{Model Architecture} How much money a user is planning to use for investment is not only related to the user's investment intention and investment habits but also directly related to the user's available capital.  To tackle these challenges, on one hand, we use the widely used self-attention mechanism\cite{vaswani2017attention} to capture the users' investment habits based on their historical behavior data, i.e. click funds, purchase, sell and add to favorites, in the sequential manner. On the other hand, we construct a consumption graph $G = (\mathcal{V}, \mathcal{E})$ to better estimate users' post-conversion purchase amount. The consumption graph is a bipartite graph, the nodes in the graph consist of user nodes $\mathcal{V}_U$ and merchant nodes $\mathcal{V}_M$. There is an edge between a user $u$ and a merchant $m$ if $u$ has paid to $m$.  Graph neural networks (GNNs) are proven to be effective for extracting information from graph data and have been widely used \cite{kipf2016semi,hamilton2017inductive,hong2020attention}. The user's payment behavior to different merchants contains valuable information and implies the user's available capital in a way. In addition, applying the two-layer GNN model to the user-merchant-user graph can help the model capture the information of a target user through the attributes of users who have similar consumption behavior patterns. Without loss of generality, we utilize graph attention layer \cite{DBLP:journals/corr/abs-1710-10903} to help ESDPM focus on the important information of consumption graph. 

The left part of Figure \ref{model} presents the architecture of ESDPM.   It firstly generates the user embedding $\mathbf{e_u}$ and fund embedding $\mathbf{e_f}$ via concatenating the feature embeddings of the user and fund respectively. The user behavior embedding $\mathbf{e_b}$ and consumption embedding $\mathbf{e_c}$ are then generated by the self-attention layer and graph attention layer. Finally, we concatenate them into a vector as input:
\begin{equation}
	\begin{aligned}
		\mathbf{e} = [\mathbf{e_u} \parallel \mathbf{e_f} \parallel	\mathbf{e_b} \parallel	\mathbf{ e_c}]	
		\label{input embedding} 
	\end{aligned}
\end{equation}
After that, ESDPM feeds the generated input vector to the subsequent multi-task learning network. Here we adapt the structure of MMoE \cite{ma2018modeling} to capture the relationships between tasks. The post-click conversion rate is learned as a hidden variable as ESMM\cite{ma2018entire}. We can obtain the ETV according to the output of ESDPM and the equation (\ref{final etv}). 

\subsubsection{Discussion}
For plain multi-task learning models including binary classification tasks and continuous values regression tasks, cross-entropy loss and MSE loss remain dominant presence\cite{liu2018joint,gutelman2020efficient}, and the final loss is obtained by simply adding the losses of the above mentioned two tasks. This form of multi-task loss function does not consider classification and regression tasks in a unified sample probability space but essentially models these two tasks separately. Besides, it ignores the variance differences of the samples' regression label distributions and is sensitive to outliers during the training process. Therefore, the performance of MTL models with such a loss function might be limited. 

ESJ loss divides the unconverted samples into two situations, i) no intention to convert ($\mathcal{R}_2$), ii) willing to convert but the investment amount is 0  ($\mathcal{R}_3$). Through such a reasonable distinction of samples,  the classification task and regression task are unified in the entire sample space and make the model perform supervised regression training for the observed negative samples. In addition, ESJ formulates the probability of each sample based on the assumption that the transaction value labels are heteroscedastic in terms of different samples. The formulation of ZILN loss shows that it uses the cross-entropy loss for classification tasks training on the entire sample space while the regression loss is only used for positive samples. Since ZILN loss has no regression supervision for the observed negative samples and needs to meet the implicit condition of $P(j \in \mathcal{R}_3) = 0$, it is easy for the model to erroneously overestimate the $\mu_i$ of the logarithmic purchase amount distribution for samples in the inference space.

\subsection{Large Scale Customer Allocation Solution}

With the estimated ETV as the constraint,  the remaining task of the EOCA framework is to solve the allocation problem described in section \ref{section2}. Although there are some commercial tools for solving constrained integer programming problems, they fail to solve large-scale allocation problems in a reasonable time \cite{ji2021large}. Moreover, the strict risk constraint and strict constraint that the number of allocated users for each segment must exact equal to the demand require a special design allocation algorithm for large scale customer allocation in FinTech platforms. 
\begin{algorithm}[!t]
	\caption{Heuristic Optimization Algorithm}\label{algorithm}
	\KwIn{ETV;  
		user risk tolerance: T = \{$t_i$|$\forall i, {u_i\in U} $\};  fund risk level: R =\{$r_j$\}; segments size demend: D = \{$d_j$\}; fund type number $k$;
	}
	\KwOut{$\mathbf{X}$}
	
	$ETV_{i,j} \leftarrow 0 , \forall i\forall j  r_j > t_i$\;
	\For{j = 1 to $k$} {$\alpha_j \leftarrow  \frac{ \sum_{i} ETV_{i,j}}{d_j} $}
	\For{i = 1 to $|U|$}{
		Sort $\{f_j| \forall j\leq k \}$ in descending order based on $ETV_{i,j}$\;
		Select the top-3 highest scored funds $a,b,c$\;
		$h_i \leftarrow  \frac{\alpha_a + \alpha_b}{2} \times (2\times ETV_{i,a} - ETV_{i,b} - ETV_{i,c}   );$
	}
	Sort $\{u_i| \forall i\leq|U|\}$ in descending order based on $\{h_i\}$\;
	\While{$i \leq |U| $}{
		Select the highest scored fund $a$ of $u_i$, $X_{ia} \leftarrow 1$ \;
		$U \leftarrow U  \backslash  u_i$ \; 
		$d_a \leftarrow d_a - 1$ \;
		\If{$d_a$ = 0}{$\forall i, ETV_{i,a} \leftarrow 0 $\;
			Execute from line 2 to line 10;
		}
		
	}
\end{algorithm}
\subsubsection{Description of our proposed allocation algorithm}
We summarize the proposed heuristic optimization algorithm (HA) in Algorithm \ref{algorithm}. In Line 1, we set $ETV_{i,j} $ to 0 if the risk tolerance level $t_i$ of user $u_i$ is lower than risk level $r_j$ of fund $f_j$. We introduce the factor $\alpha_j$ to estimate the satisfaction speed of the $f_j$ segment during the allocation process. If $\alpha_j$ of the fund j is larger than that of other funds, it implies that the $f_j$ segments may reach the allocated user number limit sooner, and other remaining users cannot be allocated to $f_j$ at that time. Based on $\alpha$ and ETV, the algorithm calculates the heuristic score $h$ for each user in Line 5 to Line 9. Firstly, for the user $u_i$, sort all funds in descending order of ETV to get the top 3 fund types $ a,b,c$. Secondly, respectively calculate the score difference between $ETV_{i,a}$ and $ETV_{i,b}$ as well as the score difference between $ETV_{i,a}$ and $ETV_{i,c}$, get $\Delta_1 $ and $\Delta_2$, and finally get $ h_{i} = \frac{\alpha_a + \alpha_b}{2}(\Delta_1 + \Delta_2)$. $h_{i}$ measures the decrease of objective function that may result if the user $i$ is not allocated to the current highest scored fund in time. In Line 10, the algorithm sorts all users in descending order according to the calculated $h$. The part from Line 11 to Line 19 executes the allocation process, the user with the highest $h_i$ is allocated to the fund $a$ with highest ETV refer to $u_i$ and remove $u$ from the unallocated user set $U$. If the $d_a$ is satisfied, the algorithm sets the $ETV_{k,a}$ to 0 for all remaining users and recalculates $\alpha$ and $h$. Repeat the loop until all users are allocated. 
\subsubsection{Discussion}
The higher the heuristic score $h_i$ for a user $u_i$ is, the more prominent the user’s preference for a certain promotional fund. If the allocation strategy fails to allocate $u_i$ to the fund that obtains the highest ETV regard to $u_i$, it probably leads to a more obvious loss for the objective value. In other words, the HA strategy tends to conduct delayed allocation for the users who have similar preferences for all $k$ promotional funds. Besides, the algorithm ensures that exact $d_j$ users are allocated to each promotional fund $j$ and users will not be allocated to the fund with a higher risk level than the user's risk tolerance. 
Although recent researches used solvers based on Lagrangian methods to achieve near-optimal solutions for the constrained optimization problem in practical applications\cite{zhang2020request,li2020spending}, these existing approaches can hardly ensure no constraints will be violated after allocation.

\section{Experiments}\label{section5}
In this section, we present the offline experiments on two real-world financial datasets and online experiments in actual delivery after customer allocation for promotional recommendations. 

\subsection{Offline Experiments}
For the offline experiments, we use two real-world datasets collected from LiCaiTong platform, which are summarized as follows:

\textbf{LCT-D}:  We collected samples from historical delivery in 2021/04/24-2021/06/01 as training (90\%) and validation (10\%) samples, which involves 8 different fund types ($k=8)$ and 4,751,065 users. We evaluate the performance of models on the samples in 2021/06/02 - 2021/06/08. The testing set contains 3,459,177 users.

\textbf{LCT-W}:  The dataset is sampled from the whole log of LiCaiTong in 2021/08/14-2021/09/20 as training (90\%) and validation (10\%) samples, which involves 4 different types of funds ($k=4$) and 4,287,959 users. We evaluate the performance of models on the samples in 2021/09/21. The testing set contains 688,171 users. Different from \textbf{LCT-D}, this dataset contains samples on entire platforms rather than being limited to samples from historical delivery.

In order to demonstrate the effectiveness of our proposed ESDPM for practical industrial application,  we conduct extensive experiments on the aforementioned datasets. AUC and logloss are used as the metrics for conversion prediction task evaluation. MAE and MSE are used as evaluation metrics for ETV prediction task evaluation as most other works that involve continueous value prediction task\cite{wang2021multi,abdellah2019iot,chiang2018personalized}. We compare ESDPM with several widely used methods and a variant model of ESDPM without the consumption graph, which are described as follows:

\begin{itemize}
	\item $\mathbf{DNN}$: Vanilla DNNs cannot handle multi-task learning data. Therefore, we trained two DNN models for click\&conversion rate prediction task and logarithmic purchase value estimation task, denoted as $\mathbf{DNN\mbox{-}C}$ and $\mathbf{DNN\mbox{-}V}$ respectively.
	
	\item $\mathbf{Shared\mbox{-}Bottom}$: It's a commonly used structure for multi-task learning. We leverage 3 shared bottom layers and two tower networks with 2 layers for classification and regression tasks respectively. 
	
	\item $\bf{MMoE}$ \cite{ma2018modeling}:  MMoE has been widely accepted as an effective multi-task learning model and deployed in the actual industrial application\cite{zhao2019recommending}, it leverages multiple expert networks and the gating mechanism to capture the relationship between different tasks. Note that MMoE does not learn the hidden variable post-click CVR, but directly predicts the probability of a user clicking and converting. 
	
	\item $\bf{MTL\mbox{-}ZILN}$: It uses exactly the same model structure as the ESDPM for multi-task learning while the ZILN loss \cite{wang2019deep} is used to train the model.

	\item $\bf{MTL\mbox{-}MSE}$: The structure of this model is similar to ESDPM, but it does not have the tower network of $\sigma$, and the regression task directly fits the logarithmic purchase amount with MSE loss function as well as MMoE. pCTCVR is obtained by multiplying the learned pCTCVR and pCVR.

	\item $\bf{ESMM}$:  It regards the post-click conversion rate as a hidden variable and estimates the pCTCVR as the product of pCTR and pCVR. The plain ESMM network only uses the shared embedding layer as the bottom structure, and we replace the network structure with the MMoE structure. ESMM is a specially designed model for conversion rate prediction and thus there are no towers for the regression task.

	\item $\bf{ESDPM\mbox{-}NCG}$: It is a viariant model of ESDPM, which removes the consumption graph from the input information as well as the coresspoding graph attention layers.

\end{itemize}
	
\begin{table}[!t]
	\caption{Comparision of models on \textbf{LCT-D}}
	\centering
	\label{tab:commands}
	\label{res_1}
	\begin{tabular}{ccccl}
		
		\toprule
		Model & AUC & logloss & MSE & MAE  \\
		\midrule
		ESMM & 0.8303 & 0.0243 & -& \\
		DNN$\mbox{-}C$ & 0.8237& 0.0233& -& \\
		DNN$\mbox{-}V$ & - & - & 0.0386 & 0.0413\\
		Shared$\mbox{-}$Bottom & 0.8120& 0.0239 & 0.0367 & 0.0323  \\
		
		MMoE & 0.8205 &0.0258 & 0.0381 & 0.0392 \\
		MTL$\mbox{-}$ZILN& 0.8287 &0.0226 & 0.0386 & 0.0426 \\
		MTL$\mbox{-}$MSE& 0.8276 &0.0236 & 0.0382 & 0.0353 \\
		ESDPM$\mbox{-}$NCG & \textbf{0.8342}& 0.0193& 0.0365 & 0.0265  \\
		$\mathbf{ESDPM}$& 0.8327 & \textbf{0.0185}& \textbf{0.0362} &\textbf{0.0227}  \\
		\bottomrule
	\end{tabular}
\end{table}

\begin{table}[!t]
	\centering
	\caption{Comparision of models on \textbf{LCT-W}}
	\label{res_2}
	\label{tab:commands}
	\begin{tabular}{ccccl}
		\toprule
		Model & AUC & logloss & MSE & MAE  \\
		\midrule
		ESMM & 0.9270 & 0.0887 & -& \\
		DNN$\mbox{-}C$ & 0.9040& 0.1013& -& \\
		DNN$\mbox{-}V$ & - & - & 0.4248 & 0.1369\\
		Shared$\mbox{-}$Bottom & 0.9208& 0.0920 & 0.4265 & 0.1404  \\
		
		MMoE & 0.9192& 0.0917 & 0.4236 & 0.1436 \\
		MTL$\mbox{-}$ZILN& 0.9228&0.0927 & 1.7885 & 0.5909 \\
		MTL$\mbox{-}$MSE& 0.9213 &0.0907 & 0.4262 & 0.1446 \\
		ESDPM$\mbox{-}$NCG &0.9283& 0.0900& 0.4206 & \textbf{0.1251}  \\
		$\mathbf{ESDPM}$& \textbf{0.9313}&\textbf{ 0.0880}&\textbf{ 0.4203} & 0.1326  \\
		\bottomrule
	\end{tabular}
\end{table}

For a fair comparison, all models use the same sequence of user historical behavior and generate user behavior embedding with the self-attention mechanism for each dataset. The behavior embedding size is set to be 128 for all models.  All models are optimized with Adam\cite{kingma2014adam}. We adopt the early stopping strategy on the loss of validation set and fix the batch size to be 512 for all cases. The learning rate, dropout probability, and other hyper-parameters are tuned for each model according to the performance on the validation set with a grid search strategy.  

\subsubsection{Effectiveness of ESDPM}The experimental results on the two real-world datasets are reported in Table \ref{res_1} and Table \ref{res_2} respectively. The results illustrate that our proposed model achieves the best performances in both conversion rate prediction tasks and expected transaction value prediction tasks. In addition, the evaluation results of the entire sample space show that the model trained with ZILN has a larger error compared with models trained with MSE for ETV estimation. It indicates that the sample selection bias caused by ignoring negative samples in the ETV estimation task leads to unsatisfactory performance on the entire inference sample space. 

\begin{figure}[!t]
	\centering
	\includegraphics[width=0.85\linewidth]{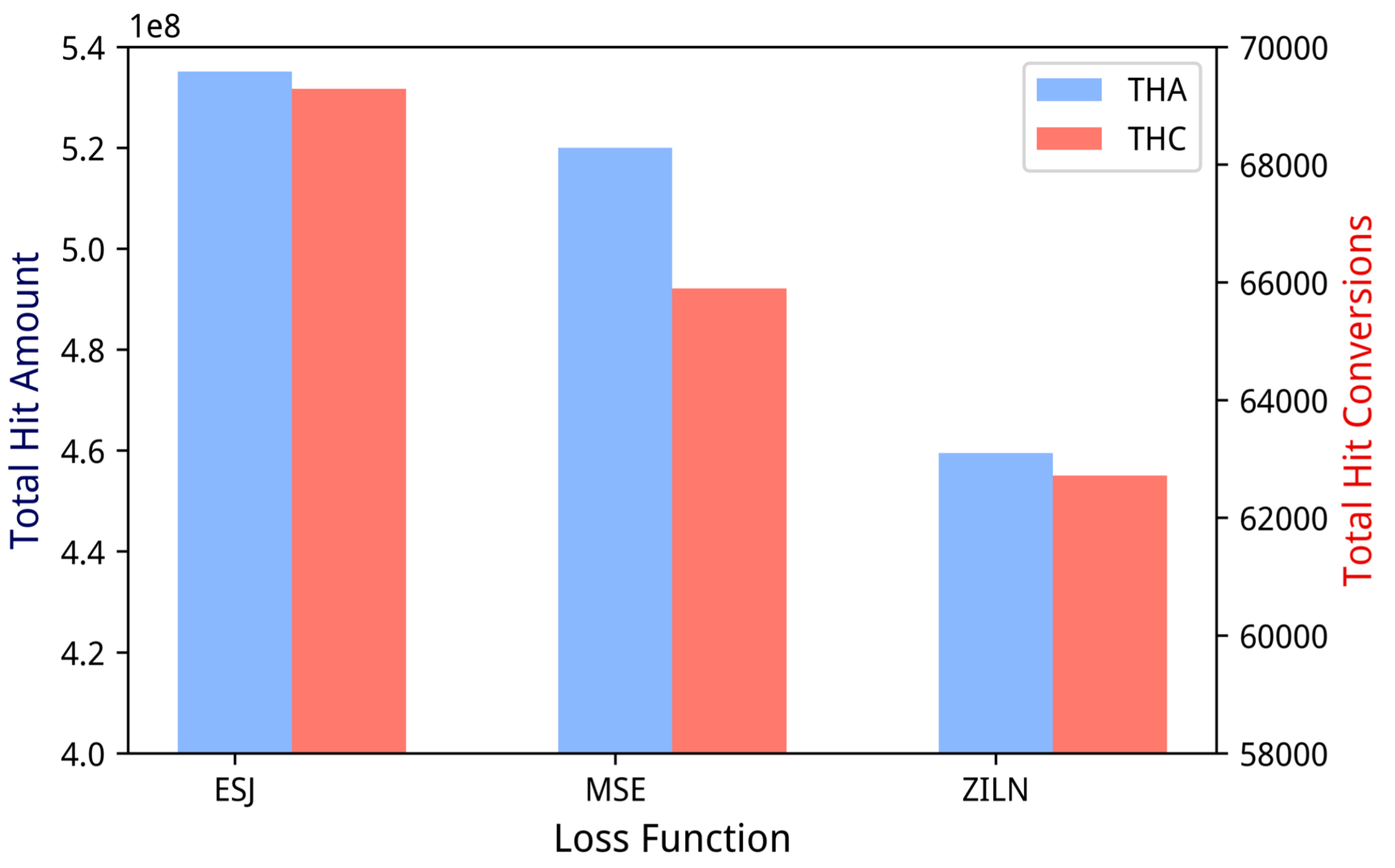}
	\caption{Performance of different loss functions.}
	\label{backtest}
\end{figure}

It is worth noting that the MTL$\mbox{-}$MSE and MTL$\mbox{-}$ZILN have the same bottom structure and paradigm for conversion rate prediction as ESDPM$\mbox{-}$NCG. To further illustrate that our proposed ESJ loss function is effective for customer allocation in financial scenarios, we conduct a simulation experiment on $ \bf{LCT}-W$ comparing models with different loss functions.  To be specific, we adopt the models trained with three types of loss function mentioned above to predict the preference score of testing samples respectively, each user $u_i$ is assigned to a group $G_j$ if $u_i$'s ETV for fund type $f_j$ is the highest among the four types of funds.  Two metrics are used in this simulating experiments:

\textbf{Total Hit Conversions}. The total hit conversions (THC) measures how many users invest in the fund with the highest predicted ETV. It is formulated as:
\begin{equation}
	\begin{aligned}
		\mathbf{THC} = \sum_{j} \sum_{\{i| u_i \in G_j \}} {y_{i,j}}
		\label{TEC} 
	\end{aligned}
\end{equation}

\textbf{Total Hit Amount}. The total hit amount (THA) measures how much money users invest in the fund with the highest predicted ETV. It is formulated as:
\begin{equation}
	\begin{aligned}
		\mathbf{THA} = \sum_{j} \sum_{\{i| u_i \in G_j \}} {PA_{i,j}}
		\label{TEC} 
	\end{aligned}
\end{equation}
The experimental results are shown in Figure \ref{backtest}. The model trained with our proposed ESJ loss achieve the best performance in both THC and THA. The results prove that ESJ can help model better estimate users' investment preferences comparing with MSE and ZILN. It also indicates that the model trained with ESJ loss can help deliver the suitable financial product to achieve the maximum total transaction value in actual applications.

\begin{table*}
	\center
	\caption{Performance of different framworks in multiple online A/B tests}
	\label{online}
	\begin{tabular}{ccccccc}
		\toprule
		Delivery Period & Model & Allocation Strategy& CPMD & TAPMD & \%CPMD Lift    & \%TAPMD Lift \\
		\midrule
		\multirow{2}{*}{ P1} & {ESMM}& Manual & 2.69 & 14,901 & - & -\\ %end line
		\cline{2-7} %short partial horizontal lines from column 2 to column 5
		&  ESMM & HA & 2.88 & 13,492 & 7.06\% & -9.46\%\\ %first cell is occupied by the multirow
		\hline
		\hline
		
		\multirow{2}{*}{ P2} & {ESMM}& HA & 2.05 & 7,359 & - & -\\ %end line
		\cline{2-7} %short partial horizontal lines from column 2 to column 5
		&  {ESDPM$\mbox{-}$NCG} & HA & 2.55 & 10,048 & 24.39\% & 36.54\%\\ %first cell is occupied by the multirow
		\hline 
		\hline 
		\multirow{2}{*}{ P3} & {ESMM}& Manual & 3.39 & 16,608 & - & -\\ %end line
		\cline{2-7} %short partial horizontal lines from column 2 to column 5
		&  {ESDPM$\mbox{-}$NCG} & HA  & 4.29 & 25,558 & 26.54\% & 53.89\%\\ %first cell is occupied by the multirow
		\hline 
		\hline  
		\multirow{2}{*}{ P4} & {ESDPM$\mbox{-}$NCG}& HA & 5.22 & 24,636 & - & -\\ %end line
		\cline{2-7} %short partial horizontal lines from column 2 to column 5
		&  {ESDPM} & HA & 5.39 & 28,568 & 3.26\% & 15.96\%\\ %first cell is occupied by the multirow
		\hline
	\end{tabular}
\end{table*}

\subsubsection{Effectiveness of HA}The offline experiments on two-real world datasets verify the effectiveness of our proposed ESDPM model for expected transaction value prediction in financial product recommendations. The next step is to allocate customers into $k$ groups. Although the traditional integer programming solver such as branch and cut algorithm can achieve remarkable objective value for such a combinatorial optimization problem, its computation cost is unacceptable for practical industrial application that involves millions of users. A simple but applicable resource allocation strategy is to manually define the priority of the funds according to the average historical conversion rate of the funds and the experience of operation specialists. For the fund type with the highest priority $f$, sort all users in descending order according to their preference scores that predicted by the model for $f$, the top-$d_i$ users are allocated to $f$, and so on until the allocation is completed. Before our proposed heuristic algorithm (HA) is deployed, the manual allocation strategy is used for the constrained customer allocation in LiCaiTong platform. However, the manual allocation strategy leads to limited performance and highly relies on the operation specialist's experience.

\begin{figure}[h] \centering    
	\subfloat[The objective value of different allocation strategies.] {
		\label{obj}
		\includegraphics[width=0.49\columnwidth]{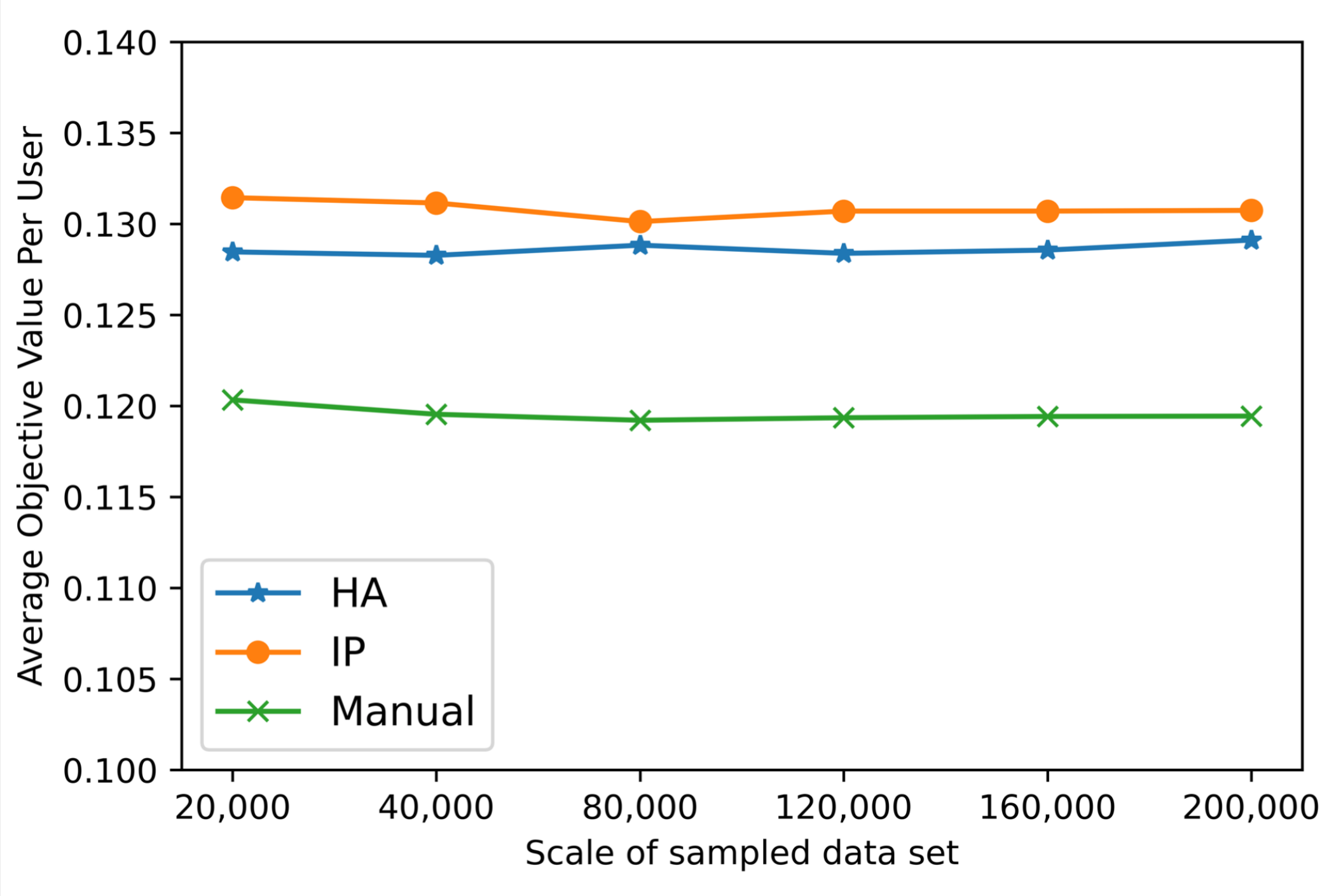}  }     
	\subfloat[The time cost of different allocation strategies. ] { 
		\label{time_cost}
		\includegraphics[width=0.49\columnwidth]{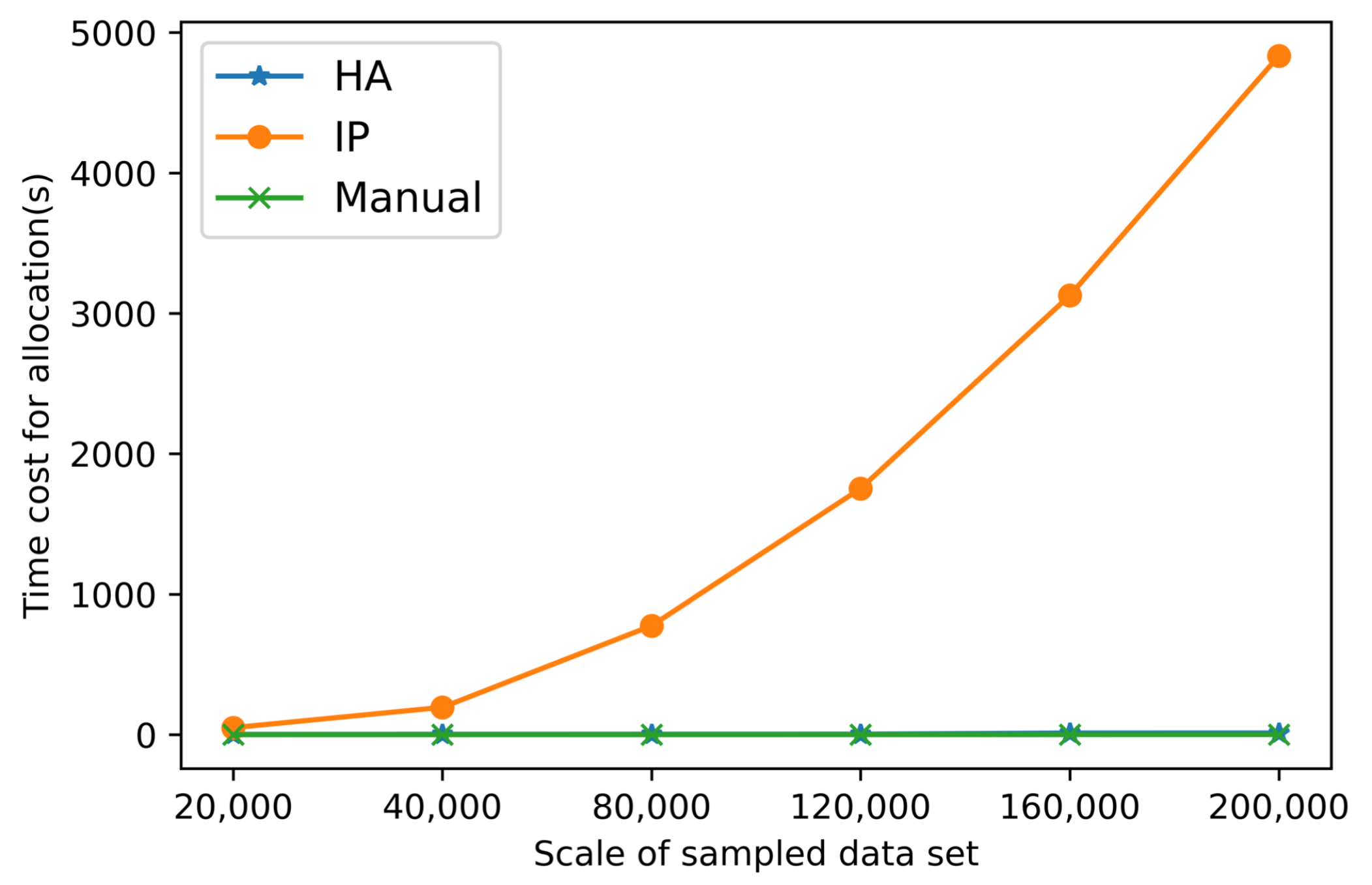}}    
	\caption{Comparision of different allocation strategies.  }     
	\label{ip_cmp}     
\end{figure}

In this experiment, we randomly sample datasets of different scales from the actual historical delivery. With the predicted ETV and delivered demand of different funds, we simulate the allocation and compare the performance of our proposed heuristic algorithm with the traditional integer programming (IP) solver and the manual allocation strategy. As shown in Figure \ref{obj}, the IP achieves the best objective. Our proposed HA allocation strategy can obtain similar performance as IP. For example, in the dataset with 200,000 users, HA achieves the 98.75\% objective score of IP. Figure \ref{time_cost} shows the time cost of different strategies. It demonstrates that the time consumption continues significantly outperform  IP in different data scales. In the large dataset with 200,000 users, the speed-up ratio achieves 416. In practical application, we need to solve the allocation problem involves in over 10 million users and the solution time for traditional IP methods is unacceptable.

\subsection{Result from online A/B testings }
We have conducted multiple online A/B testing experiments during several delivery periods to demonstrate that the methods we proposed are effective in practical industrial applications. The number of delivered users is 10 million of each period. For online A/B testing, we focus on metrics: i) Conversion Per Mille Deliveries (CPMD), ii) Transaction Amount Per Mille Deliveries (TAPMD), which are formulated as follows: 
\begin{equation}
	\begin{aligned}
		\mathbf{CPMD} =\frac{ \mathbf{\# \ conversion}}{\mathbf{\#  \ deliveries}} \times 1000
		\label{CPMD} 
	\end{aligned}
\end{equation}

\begin{equation}
	\begin{aligned}
		\mathbf{TAPMD} =\frac{ \mathbf{\sum{PA}}}{\mathbf{\#  \ deliveries}} \times 1000
		\label{TAPMD} 
	\end{aligned}
\end{equation}

In the online experiments, we compared multiple frameworks, i.e. combinations of models and allocation strategies, to illustrate the superiority of our proposed EOCA framework. The results are summarized in Table \ref{online}. It should be noted that because the market conditioens in each delivery period are different, and the fund products delivered for recommendation are also different, there is no comparison between the performances of frameworks in different two delivery periods. The major observations from the online experimental results are summarized as follows:

\begin{itemize}
	\item In P1, both two frameworks used pCTCVR predicted by ESMM as the ETV for allocation optimization, and the framework that utilizes our proposed heuristic algorithm as allocation strategy achieved an improvement in the actual online conversion rate compared to the manual allocation strategy. However, TAPCD decreased due to ignoring the expected purchase amount in the ETV estimation phase. The experimental results from $P1$ indicate that an increase in the conversion rate does not mean an increase in the total transaction amount for financial scenarios. For example, some users may have a strong intention to invest in a certain financial product, but the investment amount is limited due to concerns about risks. This shows that to achieve the optimization of sales amount, it is necessary to estimate the expected subscription amount for each user-fund pair.
	
	\item Experimental results from P2 and P3 demonstrate that our proposed EOCA framework is effective for online guarantee delivery. To be specific, both frameworks utilize HA as the allocation strategy while EOCA takes the purchase amount estimation into consideration and thus significantly improves the transaction amount by 36.54\%. In addition, the EOCA also improves the CPMD, which implies that it can better estimate the customer's purchase intention with the entire sample space loss function we proposed. In P3, we compare the EOCA framework with the original applied guaranteed delivery framework in LiCaiTong. It proves that the customer optimized segmentation framework we proposed in this paper is superior to the current practical solutions.
	
	\item In the delivery period P4, we conducted an online ablation experiment to illustrate the importance of the consumption graph. The experimental results imply the CG contains valuable information which can help ESDPM better estimate customers' purchase intention and ETV.

\end{itemize}

\section{Conclusion}\label{section6}
In this paper, we study the problem of constrained customer allocation for FinTech platforms that distribute mutual funds, which aims to maximize the total transaction value for the promotional funds. To tackle this challenging problem, we present the ETV-optimized customer allocation framework that firstly estimates the expected transaction value for each delivery and secondly allocates users to different fund types for the corresponding promotional fund to deliver. We propose an entire space multi-task joint loss function that unifies the classification task and regression task in the entire sample probability space and we present the entire space deep probabilistic model based on ESJ for ETV estimation. To the best of our knowledge, it's the first attempt to predict customers' purchase amount on FinTech platforms. In addition, we propose an effective heuristic algorithm with the predicted ETV as the constant for the allocation stage, which can achieve near-optimal solutions efficiently under multiple constraints.  The framework has been successfully deployed in the Tencent LiCaiTong platform. Offline experiments and online A/B tests demonstrate the effectiveness of EOCA.

\bibliographystyle{ACM-Reference-Format}
\bibliography{main}

\end{document}